%% file: main.tex
\documentclass[sigplan]{acmart}

\renewcommand\footnotetextcopyrightpermission[1]{}
\acmConference[2024]{2024}{}{}
\setcitestyle{acmnumeric}

\usepackage[shortlabels]{enumitem}
\usepackage{xcolor}

\newcommand{\ia}{\textit{i}}
\newcommand{\ib}{\textit{ii}}

\newcommand{\temph}[1]{\textbf{#1}}

\newcommand{\udi}[1]{\textcolor{blue}{[Udi: #1]}}

\title[The Blocklace as a CRDT]{The Blocklace: A  Byzantine-repelling and Universal \\ Conflict-free Replicated Data Type}

\author{Paulo Sérgio Almeida}
\affiliation{%
  \institution{INESC TEC \& University of Minho}
\country{Portugal}
}
\author{Ehud Shapiro}
\affiliation{%
  \institution{Weizmann Institute of Science}
  \country{Israel}\\
   \institution{London School of Economics} 
  \country{United Kingdom}
}

\mathcode`|="326A          
\newcommand\af[1]{\mathop{\hbox{$\mathsf{#1}$}}\nolimits}
\newcommand\auxfun[1]{\expandafter\newcommand\csname #1\endcsname{\af{#1}}}
\newcommand\defeq{\mathbin{\stackrel{.}{=}}}
\newcommand\union{\cup}
\newcommand\bigunion{\bigcup}
\newcommand\join{\sqcup}
\newcommand\pointed{\leftarrow}

\newcommand{\true}{\mathsf{True}}

\let\max=\relax
\auxfun{max}
\auxfun{hash}
\auxfun{signedhash}
\auxfun{node}
\auxfun{nodes}
\auxfun{dom}
\auxfun{id}
\auxfun{ids}
\auxfun{new}
\auxfun{add}
\auxfun{eqvc}
\auxfun{wf}
\auxfun{valid}
\auxfun{byz}
\auxfun{brep}
\auxfun{polog}
\auxfun{prepare}
\auxfun{effect}

\RequirePackage{amsmath, amsthm, mathtools}

\begin{abstract}

Conflict-free Replicated Data Types (CRDTs) are designed for replica convergence without global coordination or consensus. Recent work has achieved the same in a Byzantine environment, through DAG-like structures based on cryptographic hashes of content.
The blocklace is a partially-ordered generalization of the blockchain in which each block has any finite number of signed hash pointers to preceding blocks.  We show that the blocklace datatype, with the sole operation of adding a single block, is a CRDT: it is both a pure operation-based CRDT, with self-tagging; and a delta-state CRDT, under a slight generalization of the delta framework. Allowing arbitrary values as payload, the blocklace can also be seen as a universal Byzantine fault-tolerant implementation for arbitrary CRDTs, under the operation-based approach.
Current approaches only care about CRDT convergence, being equivocation-tolerant (they do not detect or prevent equivocations), allowing a Byzantine node to cause an arbitrary amount of harm by polluting the CRDT state with an unbounded number of equivocations.
We show that the blocklace can be used not only in an equivocation-tolerant way, but also so as to detect and eventually exclude Byzantine nodes, including equivocators, even under the presence of undetectable colluders. 
The blocklace CRDT protocol ensures that a Byzantine node may harm only a finite prefix of the computation.
\end{abstract}

\begin{document}


\maketitle

\input{intro}

\input{blocklace}

\input{crdt}

\input{byzantine}

\section{Cordial Dissemination}

The dual nature of a blocklace as both a pure op-based with self-tagging and a delta CRDT hints to weaker demands in terms of dissemination protocols. Block identity and their causal relationship are incorruptible and stored in the blocks themselves. Therefore, we do not need to rely on traditional causal delivery~\cite{DBLP:journals/tocs/BirmanSS91}, with external tracking of causality, which would be prone to Byzantine faults, from Byzantine nodes participating in the protocol. With self-tagging of identity and predecessors, duplicates are easily discarded, and missing predecessor blocks can be requested and waited for.  Indeed, the blocklace-based Cordial Dissemination protocol~\cite{keidar2021need,lewispye2023flash} employs the fact that a $p$-block $b$ by a correct node $p$ acknowledges all the blocks known to $p$ at the time of block creation (and, by omission, reports on all the blocks not known to $p$ at the time). Thus, when $q$ accepts the $p$-block $b$ it eventually cordially responds by sending to $p$ all the blocks known to $q$ but not to $p$, according to $b$, achieving eventual visibility in the face of Byzantine nodes.
The response may be arbitrarily lazy or even probabilistic.

\section{Conclusions}

We have shown how a blocklace can be seen as both a pure operation-based CRDT, with self-tagging, and a delta-state CRDT, under a slight generalization of the delta framework. With arbitrary values as payload, the blocklace can be seen as a universal Byzantine fault-tolerant implementation for arbitrary CRDTs, under the operation-based approach.

Contrary to current DAG-like structures that use hashes of content, a blocklace uses signed hashes from which the block creator can be known. Not only this can aid dissemination, by knowing which node is missing which blocks, but more importantly, it can be used to detect equivocations and equivocators.

Current equivocation-tolerant approaches do not detect or prevent equivocations, allowing a Byzantine node to cause an arbitrary amount of harm by polluting the CRDT state with an infinite number of equivocations. This can hinder optimizations for efficient query processing, where the blocklace is used as robust storage of knowledge, complemented by auxiliary data structures.

We have shown how Byzantine behavior can be detected and eventually excluded, limiting harm to only a finite prefix of the computation. The most challenging aspect is the possible presence of colluders: Byzantine nodes that act so as to remain undetected as such by correct nodes, but would force the incorporation of an unbounded number of blocks from an equivocator if nothing was done to prevent it.

We show a solution to achieve limited harm: given that colluders cannot be proved as such by correct nodes in an asynchronous distributed system, the protocol forces a node to acknowledge already known Byzantine nodes (including equivocators) before incorporating its blocks.

\begin{acks}
Ehud Shapiro holds The Harry Weinrebe Professorial Chair of Computer Science and Biology at  Weizmann. 
\end{acks}
\bibliographystyle{ACM-Reference-Format}
\bibliography{bib,local}

\input{appendix}

\end{document}

%% file: intro.tex
\section{Introduction}

The blockchain is a data structure in which each block except the first has a hash pointer to its predecessor.
It is constructed competitively, by nodes competing to add the next block to the blockchain.
Blockchain consensus protocols such as Nakamoto
Consensus~\cite{bitcoin} resolve the conflicts, or `forks', that arise
when two or more nodes add their own blocks to the same blockchain.  

The blocklace is a partially-ordered generalization of the blockchain in which
each block has any finite number of signed hash pointers to preceding blocks. It can be
constructed cooperatively, by nodes adding their signed blocks to the blocklace
and informing each other of new blocks. 



The description above suggests that, while the blockchain is fundamentally a
conflict-based datatype, the blocklace is not. Indeed, this paper casts the
blocklace as a Conflict-free Replicated Data Type
(CRDT)~\cite{DBLP:conf/sss/ShapiroPBZ11} and analyzes its properties and
related protocols as such.

We show that the blocklace datatype, with the sole operation of adding a
single block, is a CRDT. We show that it is both a pure operation-based CRDT~\cite{DBLP:conf/dais/BaqueroAS14} and a delta-state CRDT~\cite{DBLP:journals/jpdc/AlmeidaSB18}, under a slight generalization of
the framework.

Furthermore, even though it is a concrete datatype it is quite general as
such, given the ability to store arbitrary payloads and encoding the
causality relationship between updates. It can, therefore, serve as a universal
store, isomorphic to the PO-Log of pure operation-based CRDTs, allowing
different CRDTs to be implemented.

CRDTs are designed to allow replica convergence with no need for global
coordination or consensus. Recent
work~\cite{DBLP:journals/corr/abs-2004-00107,DBLP:conf/eurosys/Kleppmann22,DBLP:conf/sicherheit/JacobBH22}
has shown that the same can be achieved for CRDTs under Byzantine
environments, through DAG-like structures, or some variation thereof~\cite{DBLP:conf/eurosys/JacobH23} based on cryptographic hashes of
content.

As a first ingredient, these approaches achieve Byzantine fault tolerance
(BFT) assuming~\cite{DBLP:journals/corr/abs-2012-00472} (as we do) that correct nodes form a connected graph, so that
each update delivered by a correct node is forwarded along correct nodes,
and eventually delivered by all correct nodes. A second ingredient, given that
Byzantine nodes can issue malformed or semantically invalid updates, is that
each correct node must decide in the same way whether a received update is accepted or
rejected. As discussed in~\cite{DBLP:conf/eurosys/Kleppmann22}, this can be
achieved by having the decision depend on the causal past of the update
itself, regardless of concurrent updates that may be present.

A more challenging Byzantine behavior that cannot be detected solely based on the causal history of a block is \emph{equivocation}: sending different updates to different nodes, as opposed to the same update to all nodes.
Current solutions to this challenge are \emph{equivocation-tolerant}~\cite{DBLP:conf/sicherheit/JacobBH22,DBLP:conf/eurosys/JacobH23}, as they do
not detect, prevent or remedy equivocation. Messages that are not rejected as invalid but could come from an equivocating Byzantine node are simply accepted and treated as if they were concurrent operations by different
correct nodes.

While this approach achieves convergence, it allows a Byzantine node to cause an
arbitrary amount of harm by polluting the CRDT state with an unbounded number
of equivocations. These can hinder optimizations towards achieving efficient
query processing. A single Byzantine node can create states denoting a
degree of concurrency only achievable by many correct nodes.

In this paper we show how the blocklace can be used not only to recast extant
equivocation-tolerant protocols, but also to detect equivocations and eventually exclude
all Byzantine nodes, in particular equivocators. The blocklace Byzantine fault-tolerant protocol ensures that a Byzantine node can harm only a finite prefix of the computation.

In current DAG-like structures such as the Block DAG~\cite{lewenberg2015inclusive}, the identity of a block, that serves as pointer to the block, to be stored in other blocks, is a hash of the block content, with no information about which node created the block. The Block DAG was used to provide  multiple improvements and extensions of the Nakamoto Consensus protocol~\cite{sompolinsky2021phantom,sompolinsky2016spectre,sompolinsky2022dag}.

The distinguishing feature of the blocklace is that the block identity, used for pointing to other blocks, is a signed hash of the block content. The signature allows knowing which node created the block, in an unforgeable fashion, and is the basis from which we could build the equivocation detection and exclusion algorithms that we present, which are one of the main contributions. We note that simply storing the public key of a node in the block content, to identify the creator, is not Byzantine tolerant, as a Byzantine node can ``blame'' another node as the block creator.

Previous applications of the blocklace datatype include grassroots dissemination~\cite{shapiro2023grassroots,shapiro2023grassrootsBA}, grassroots social networking~\cite{shapiro2023gsn},  a payment system~\cite{lewispye2023flash}, an implementation of grassroots currencies~\cite{lewis2023grassroots},  a family of Byzantine Atomic Broadcast consensus protocols~\cite{keidar2023cordial}.


%% file: blocklace.tex
\section{The Blocklace}

\subsection{Hashes, Signatures and Identities}

We assume a distributed system with a finite set of nodes $\Pi$, each identified by a single pair of unique private and public keys. The public key $p\in \Pi$ identifies the node.

We assume a collision-resistant cryptographic hash function $\hash$, for which it is computationally infeasible to find $x$ from $\hash(x)$, and for which the probability of $\hash(x) = \hash(y)$ for $x\ne y$ is infinitely small.  Practically, SHA-256 can be used.
A node can sign some content (e.g., some hash value $h$) using its private key,
obtaining a signature $s$.  We assume that the identity (public key) of the
node can be extracted from its signature, e.g., using
ECDSA.

\subsection{Blocks}

\begin{definition}[Block]
A \emph{block} $b$ created by node $p$, or $p$-block, is a pair $(i, C)$, where 
$C$ is the \emph{block content}
and $i$,  the \emph{block identity},  is $\hash(C)$ signed by $p$.
The block content is a pair $C=(v, P)$ of an arbitrary value $v$ (the block \emph{payload}) and
a set $P$ of the block identities of the \emph{predecessors} of $b$;  if $P=\emptyset$ then $b$ is \emph{initial} (or \emph{genesis}).
\end{definition}

Thus, each block pairs a unique block identity 
with block content, which holds an arbitrary payload (e.g., defining
an operation, a list of operations or a list of transactions) and pointers to
predecessor blocks through their block identities.

Contrary to other approaches, each pointer allows not only uniquely
identifying a block, but also knowing the block provenance, i.e., its creator. For a $p$-block $b = (i, C)$ we use both $\node(b)=p$ and
$\node(i)=p$ for the block creator identity, and $\id(b)$ for the block identity, i.e., $i$.
The function $\nodes$ denotes the set of creators of a set of blocks:
\[
  \nodes(S) \defeq \{\node(b) | b \in S\},
\]
and similarly for the set of identities:
\[
  \ids(S) \defeq \{\id(b) | b \in S\}.
\]


We remark that even if two nodes $p,q \in \Pi$ hold identical blocklaces, and each of these
nodes creates a new block with the same payload and same set of predecessor
blocks, and therefore with the same hash, block identities will be
different as the hash will be signed by the different private keys of $p$ and $q$.



\subsection{Blocklaces}

A \emph{blocklace} $B$ is a set of blocks subject to some invariants described
below.

\begin{definition}[Pointed]
  A block $a = (i_a, C_a)$ is pointed from a block $b = (i_b, (v_b, P_b))$ if
  and only if $i_a \in P_b$. If block $a$ is pointed from block $b$ we write $a
  \pointed b$.
\end{definition}

\begin{definition}[Closure axiom]
A blocklace $B$ must be downward closed under the pointed
relation:
\[
  a \pointed b \land b \in B \implies a \in B.
  \tag{$\text{CLOSED}$}
\]
\end{definition}

Thus, there are no ``dangling pointers'' in a blocklace: each identity in
each set of predecessors of a block in a blocklace identifies a block in the blocklace.

Since each block is created with a unique identity (even if blocks have
identical content) and a blocklace is a set of pairs, it can also be
seen as a map (finite function) from block identity to block content.
This is notationally useful. For example, the closure axiom can also be
written as:
\[
  \forall (i, (v, P)) \in B \cdot P \subset \dom(B).
\]
i.e., the predecessors of each block are in the map domain.
Given that a blocklace $B$ is a function, $B(b)$ is the block content of the
block with id $b$. We use $B[b]$ for the block itself, i.e.,
\[
  B[b] \defeq (b, B(b)),
\]
and similarly for a set of blocks, given a set of pointers:
\[
  B[P] \defeq \{ B[b] | b \in P \}
\]

\begin{definition}[Precedes]
  Block $a$ \emph{precedes} block $b$, written $a \prec b$ iff $a$ is directly or
  transitively pointed from $b$:
\[
  \prec \defeq \pointed^+.
\]
\end{definition}
I.e., precedes is the transitive closure of pointed.
$a\preceq b$ denotes $a \prec b$ or $a=b$.
If $a \prec b$ we also say that $b$ \emph{observes} $a$.

\begin{proposition}[Acyclicity]
The relation $\prec$ is acyclic.
\end{proposition}

This follows trivially from the the fact that $\hash$ is cryptographic, which prevents compute-bound nodes, including byzantine ones, from computing cycles.
Hence $\prec$ is a strict partial order.  Moreover,
because starting from any block we can only ``go down'' a finite number of
steps, $\prec$ is also a well-founded relation.
Node that $\prec$ may hold among arbitrary blocks, not just within a blocklace.


For a relation $r$ (such as $\prec$ or $\pointed$), block $b$ and set of
blocks $S$, we use the notation $rb = \{a | a \mathrel{r} b\}$, and 
$rS = \bigunion \{rb | b \in S\}$.
In particular, we use ${\preceq}b$ for the downward closure starting at
block $b$, and similarly for ${\preceq}S$.

\begin{definition}[Virtual chain axiom]
Blocklace $B$ satisfies the virtual chains axiom for node $p\in \Pi$ if
any two different $p$-blocks $a, b \in B$ are
comparable:
\[
  \node(a) = \node(b) = p \implies a \prec b \lor b \prec a.
  \tag{$\text{CHAIN}$}
\]
\end{definition}

Even though a blocklace is partially ordered, if it satisfies
the chain axiom for some node $p$, the set of $p$-blocks form a strict total order under $\prec$. Thus, even if two $p$-blocks $a$ and $b$ are not connected by direct pointers, $a\prec b$ implies the existence of a chain from $a$ to $b$ via non-$p$ blocks.


As discussed below, the virtual chain axiom always holds for correct nodes in any blocklace; nodes that violate it are Byzantine equivocators.


\subsection{Block creation and dissemination}

In a blocklace-based distributed system each node $p$ maintains two local sets of blocks:
A blocklace $B$, a closed set of blocks satisfying ${\preceq}B = B$, that includes all blocks already created by $p$ as well as blocks accepted from other agents, and a buffer---a set of buffered or delayed blocks $D$ that were received but not yet accepted. In particular, a received block that points to blocks not yet in $B$ is buffered, as its acceptance into $B$ would violate the closure axiom.

Given an arbitrary payload $v$, the operation $\add(v)$ executed by a node $p$ with blocklace $B$ 
adds to $B$ a new $p$-block with payload $v$, resulting in $B'$, as follows: 
\[
  B' = B \union \{\af{new}_p(B, v)\},
\]
where
\[
\begin{split}
  \af{new}_p(B, v) & \defeq (i, (v, P)), \text{ with} \\
  P &= \ids(\max_\prec(B)), \\
    i &= \signedhash((v, P), k_p)
\end{split}
\]
were $\max_\prec(B)$ is the set of maximal blocks in $B$ under the $\prec$
relation, i.e., the blocks which are not pointed from any other block in $B$,
and $k_p$ is the private key of node $p$.

Note that the virtual chain axiom holds for correct nodes since they create blocks in sequence and include each new block in their blocklace.
We assume an underlying blocklace dissemination protocol (e.g. Cordial Dissemination~\cite{keidar2023cordial,shapiro2023grassroots,shapiro2023grassrootsBA}) that 
satisfies the following Dissemination Axiom:

\begin{definition}[Dissemination Axiom]\label{definition:dissemination-axiom}
If the blocklace of a correct node $p$ has a block $b$ then every correct node eventually receives $b$.
\end{definition}
Note that the Axiom in fact implies that eventually ${\preceq}b$ is received by all correct nodes, since a correct node $q$ cannot incorporate $b$ in its blocklace without also incorporating its closure ${\preceq}b$.


%% file: crdt.tex
\section{The Blocklace as a Universal CRDT}

\subsection{The Blocklace as a universal replicated data type}

A blocklace is a concrete linked data structure, suitable for use in a
distributed setting with replication. Each participating node produces blocks,
which become linked to direct predecessors, and blocks are propagated aiming for
replica convergence. It can be seen as a specific replicated data type, with
the data type state being a blocklace, and with a single update operation  $\add(v)$ specified above.

As discussed below,  the blocklace with its single $\add$
operation can be viewed both as a pure operation-based CRDT and as a delta-state CRDT.
Having a single update operation may seem specific and restrictive, but the interlinked structure and the arbitrary payloads allow the blocklace to be employed as a canonical storage of a universal data
type that can be used to encode any specific data types.


\subsection{The Blocklace as as pure operation-based CRDT}

Operation-based CRDTs (op-based for short) propagate ``operations'' to other
replicas\cite{DBLP:conf/sss/ShapiroPBZ11}. In fact, they propagate the result of processing operations through
a \emph{prepare} function, which looks at the operation and the current state,
and can return an arbitrary value, to be sent in a message. Upon delivery at
each replica, an \emph{effect} function takes the state and message and
produces the next state.
A blocklace can be seen as an operation-based CRDT. For the single operation $\add$ with parameter $v$, prepare is defined as:
\[
  \prepare([\add, v], B) = \new_i(B, v),
\]
with the result being a new block, to be broadcast to other replicas.
If causal broadcast is used, as usual for op-based, effect can be defined
simply as:
\[
  \effect(b, B) = B \union \{b\}.
\]

Pure op-based CRDTs~\cite{DBLP:conf/dais/BaqueroAS14} are a special
kind of op-based CRDTs, restricted to the essence of ``send only operations'',
not arbitrary values (including the full state itself, which is possible
in the general op-based framework). Originally pure-op is defined with:
\[
\prepare(o, s) = o.
\]
To allow general CRDTs to be defined, with semantics based on the
partial order of operations, a \emph{Tagged Causal Broadcast (TCB)} middleware
is used. TCB associates a partially ordered timestamp $t$ to each message (e.g., a
version vector~\cite{DBLP:journals/tse/ParkerPRSWWCEKK83}).
The state (before optimizations) can be a PO-Log: a partially ordered log of
operations, represented as a map from timestamps to operations.
Effect simply adds an entry $(t,o)$ delivered by TCB to the PO-Log:
\[
  \effect(o, t, s) =  s \cup \{(t,o)\}.
\]

If we consider the blocklace, prepare is not doing arbitrary processing of the
operation under the current state, but simply encoding causality in the value.
This means that, instead of having to delegate to a TCB middleware, the
blocklace self-encodes causality in the block that is the result of prepare. This allows any
at-least-once dissemination mechanism, together with checking that predecessors are
already included. So, when a block $(i, (v, P))$ is received at a node holding
blocklace $B$, the node checks that:
\[
  i \not\in \dom(B) \land P \subseteq \dom(B).
\]

A blocklace $B$ is then isomorphic to a PO-Log from the pure-op framework, thus is also a pure-op CRDT.

\subsection{The Blocklace as a delta-state CRDT}

A delta-state CRDT~\cite{DBLP:journals/jpdc/AlmeidaSB18} is a state-based CRDT
in which delta-mutators return not the next state but a delta-state (delta for
short), to be joined with the current state:
\[
X' = X \join m^\delta(X).
\]

Deltas are elements of the state lattice; they can be joined into
delta-groups, to be sent in messages. A received delta-group can be joined
into the current state:
\[
X' = X \join D.
\]

The standard delta-CRDT framework assumes that joining any delta-group
to the current state produces a valid state; i.e., received delta-groups can
be joined unconditionally into the current state, allowing them to be
propagated by many different algorithms. There is only need to join
delta-groups conditionally when ensuring causal consistency, by the so
called \emph{causal delta-merging condition}.

A blocklace can be considered as a delta-CRDT by a slight generalization of
the delta-CRDT framework.
The generalization needs to distinguish a \emph{support lattice} (using
lattice here as shorthand for join-semilattice) and a \emph{state lattice}.
The generalization is:
\begin{itemize}[leftmargin=*,topsep=0pt]
  \item the state lattice is a sublattice of the support lattice;
  \item a delta-mutator produces an element of the support lattice;
  \item join is defined in the support lattice;
  \item joining deltas produces an element of the support lattice;
  \item delta-merging condition: delta-groups can be joined into local state only if the result belongs to the state lattice.
\end{itemize}
In the case of the blocklace:
\begin{itemize}[leftmargin=*,topsep=0pt]
  \item support lattice: powerset of blocks;
  \item state lattice: lattice of sets of blocks closed under $\prec$.
\end{itemize}

The join is simply set union (of sets of blocks). The delta-merging condition
amounts to checking if the union of the  state and the
delta-group produces a downward closed set of blocks under $\prec$.
Given local state $B$ (which is downward closed) and delta group $D$, the
condition becomes:
\[
  {\preceq} D \subseteq B \union D
\]
So, instead of allowing the unconditional join of delta-groups into the state, the use of conditional delta-merging ensures closure, as required for a blocklace, and also ensures causal
consistency. All the useful properties of delta-CRDTs remain:
Possibility of small messages, tolerance of duplicates through the idempotency of
join, and tolerance to message loss by allowing retransmission of the same or
newer delta-groups that subsume the lost message.

%% file: byzantine.tex
\section{The Blocklace and Byzantine Fault Tolerance}

Classic approaches to tolerating Byzantine faults, mostly aiming for strong
consistency, depend on global coordination or correct node majorities.
As CRDTs can evolve without global coordination or consensus, approaches to
tolerating Byzantine faults in the  blocklace as a CRDT should also be
free from global coordination.

Highly-available eventually consistent systems aim to satisfy the following criteria for correct
nodes: operation termination, eventual visibility~\cite{DBLP:journals/ftpl/Burckhardt14}, and convergence.\footnote{Strong convergence in~\cite{DBLP:conf/sss/ShapiroPBZ11}.} Termination
is trivial, by making operations use the local state without waiting for messages and without depending on other nodes or network condition. Regarding the other two:
\begin{itemize}[leftmargin=*,topsep=0pt]
  \item Eventual visibility is ensured by making each operation be delivered
    by all nodes (in op-based) or by incorporating the operation into the
    current state and propagating states, which makes many operations visible
    at once (state-based).
  \item Convergence, by making nodes that have the same set of visible
    operations have equivalent states. This is ensured internally by CRDTs.
    In op-based, by designing \emph{prepare} and \emph{effect} such that the effect of concurrently issued operations commutes;
    in state-based, by the state being a join-semilattice.
\end{itemize}

In Byzantine scenarios, a Byzantine node can produce corrupt or semantically
invalid updates, and corrupt or drop messages. More interestingly,
they can perform \emph{equivocation}: sending different updates to different
nodes, when a correct node would send the same update to all.

Recent approaches to Byzantine fault-tolerance for replicated data types in
eventual consistent settings use data structures such as a
Merkle-DAG~\cite{DBLP:journals/corr/abs-2004-00107},
Hash Graph~\cite{DBLP:conf/eurosys/Kleppmann22} or
Matrix Event Graph~\cite{DBLP:conf/sicherheit/JacobBH22}. Regardless of
minor differences, all use cryptographic hash functions, which prevent
Byzantine nodes from corrupting messages without being detected. These
structures also encode causality information in an incorruptible way, not
depending on external tagging by a delivery middleware.
These approaches achieve:
\begin{itemize}[leftmargin=*,topsep=0pt]
  \item Eventual visibility: Assuming that correct nodes form a connected
    graph, which allows any message received by a correct node to be forwarded
    to any other correct node along a path involving only correct nodes. This
    includes updates originated at Byzantine nodes. This overcomes omissions
    or message dropping from Byzantine nodes.
  \item Convergence: Adopting an operation-based approach, delivering messages
    respecting the incorruptible causality order, and having correct
    node similarly reject corrupt messages and accept semantically
    valid ones.
\end{itemize}

Extant approaches no not attempt to detect equivocations. Messages from Byzantine equivocators that are not rejected as invalid are simply accepted and treated as if they were concurrent operations from different correct nodes. These CRDTs are called \emph{equivocation-tolerant}~\cite{DBLP:conf/sicherheit/JacobBH22}, as they do
not detect, prevent or remedy equivocation.

As a blocklace is an incorruptible representation of information that records causality, it can be used
to realize the same equivocation-tolerant protocol to achieve eventual visibility and
convergence in a highly available system. We discuss below block acceptance
in a blocklace in terms of well-formedness and validity.

These DAG-like structures, including the blocklace, can be used as a secure
PO-Log; as a PO-Log, they warrant optimizations for querying or compact long
term storage of information which has become stable.
Efficient implementations of CRDTs, such as \emph{causal
CRDTs}~\cite{DBLP:journals/jpdc/AlmeidaSB18} use concepts such as
\emph{dots}~\cite{DBLP:journals/corr/abs-1011-5808} (pairs of node-id and
sequence number) as unique identifiers, or maps from node identifiers to
sequence numbers for causal contexts. It is important to
stress that a dot is not merely a unique identifier, but a convenient form of
unique identifier that allows causally closed sets to be compactly represented
as version-vectors, which allow trivial membership testing. On the contrary,
sets of randomly generated numbers or sets of signed-hashes (in the case of a
blocklace) do not allow such a compact representation.

Current approaches allow any Byzantine node
to cause an arbitrary amount of harm by polluting the CRDT state with equivocations.
In a system with $n$ nodes,  a single Byzantine node  can create
states denoting a degree of concurrency only achievable by much
more than $n$ correct nodes. The high degree of the underlying DAG needed to
achieve convergence will prevent optimized structures for processing queries.
It is, therefore, desirable to not only achieve convergence under Byzantine
behavior, but also to prevent, as far as possible, Byzantine nodes from
polluting the blocklace in a way that hinders such optimizations and degrades
performance.



So far we have shown that the blocklace can embody extant approaches and methods of CRDT, including equivocation tolerance. Next we show how the blocklace allows going beyond the state of the art,
by excluding all Byzantine nodes, including equivocators, during a finite prefix of the computation, leaving a potentially-infinite suffix of the computation Byzantine-free.
This is achieved by all correct nodes eventually identifying all Byzantine acts and eventually excluding any new blocks created by Byzantine nodes.  Thus the harm a Byzantine node may cause is limited to a finite prefix of the computation.

\subsection{Block well-formedness and validity}

If a block is malformed beyond recognition, e.g., the signed hash forming
its identity does not match its content, it is trivially rejected. But in this
case the culprit cannot be identified, as any node could concatenate a signed hash by another node to an arbitrary content and send it to others (and we do not assume that received messages allow identifying their sender).

Correct nodes create a new block in $B$ using the identities of the set of
maximals in $B$ as predecessors. By definition, all the elements in this set
are incomparable, forming an antichain. If that is not the case, the creator
is Byzantine.

\begin{definition}[Well-formed blocks]
A block $b$ with set of predecessors $P$ in blocklace $B$ is well-formed,
written $\wf(b, B)$ if, in addition to it having a valid signed hash, matching
the block content, all pointers in $P$ point to incomparable blocks in $B$:
\[
  \neg \exists a, b \in P \cdot B[a] \prec B[b].
\]
We also write $\wf(b)$ equivalently to $\wf(b, {\prec}b).$
\end{definition}

The second kind of Byzantine behavior, when the blocklace is used for some
data abstraction such as a CRDT, is creating a block that breaks state invariants. Some
examples~\cite{DBLP:conf/eurosys/Kleppmann22} are: reusing an identifier when
creating a supposedly unique one; referencing a nonexistent identifier, which
should exist; referencing identifiers that denote an insert position in a list
in the wrong order.


For a given abstraction, a predicate $\valid$ takes a value and a causal
past to decide weather the value could have been
produced by a correct node given that causal past. As a baseline,
considering the universal usage of the blocklace with unconstrained values,
we can assume $\valid(v, B) = \true$.
Specific abstractions (CRDTs) can redefine the $\valid$ predicate to impose constraints
on the possible values produced by correct nodes.

Given a well-formed block $b$, with predecessors $P$, received at a node holding blocklace $B$, this predicate is used to decide block acceptance. The
block can only be incorporated in blocklace $B$ if, in addition to its
predecessors being already present, i.e., $P \subseteq \dom(B)$, the block is valid: $\valid(v,{\prec}b)$, using the definition of $\valid$ provided by the data type.
Validity depends only on the value $v$ and the causal context ${\prec}b$ when $b$ was created. For notational brevity, given block $b = (i, (v, P))$, we also overload
$\valid(b) \defeq \valid(v, {\prec}b)$.

There are two design options regarding the behavior of a node upon the receipt of
a correctly-signed $p$-block that is not well-formed or not valid: 
\begin{enumerate}[leftmargin=*,topsep=0pt]
    \item Discarded the block.
    \item Accept the block into the blocklace,  resulting in all nodes eventually knowing that $p$ is Byzantine.
\end{enumerate}

The first choice only keeps valid well-formed blocks in the blocklace; for any block, all preceding blocks are part of the CRDT state.
We adopt the second option, as it ensures that in an infinite computation the harm a Byzantine node may cause is finite.  It requires that the payload of 
invalid blocks be ignored from the perspective of the date type.



\subsection{Detecting equivocations}

The third and more challenging Byzantine behavior is equivocation.
DAG-based approaches that maintain the DAG
structure~\cite{DBLP:journals/corr/abs-2004-00107,DBLP:conf/eurosys/Kleppmann22,DBLP:conf/sicherheit/JacobBH22}
with anonymous updates, do not associate an update with the node that created it and
do not have the notion of a virtual chain. Therefore they cannot identify equivocations.
On the other hand, in a blocklace the identity of the creator of a block (identified by its public key) is incorporated in the block, and therefore equivocators can be detected and identified.

\begin{definition}[Equivocation, Equivocator]
An \emph{equivocation} by a node $p$ in a blocklace $B$ is a pair of different $p$-blocks $a, b \in B$
 that are incomparable under $\prec$, i.e.:
\[
  \node(a) = \node(b) = p \land a \parallel b,
  \tag{\text{EQUIV}}
\]
where $a \parallel b$ is a shorthand for $a\not\prec b \wedge b\not\prec a$. In such a case $p$ is an \emph{equivocator} in $B$.
\end{definition}

The presence of a pair of incomparable $p$-blocks in a
blocklace proves that $p$ is an equivocator. The set of equivocators in blocklace $B$ is written
$\eqvc(B)$.
\[
  \eqvc(B) \defeq \{ p | \exists a, b \in B \cdot \node(a) =
  \node(b) = p \land a \parallel b \}.
\]

We remark that a node $p$ can belong to $\eqvc(B)$ even if no block in
$B$ observes an equivocation by $p$; the existence of a pair
of incomparable $p$-blocks in $B$ is enough.

\subsection{Byzantine nodes and the data abstraction}

According to our protocol design, the blocklace stores not only valid blocks but also blocks that constitute a proof that their creator is Byzantine. Thus only the valid subset of the blocks will be considered as state by the replicated abstraction (the CRDT). To achieve convergence, different nodes must decide in the same way whether a given block is part of the CRDT state. Thus 
the following predicate identifies whether a node $p$ is Byzantine given blocklace $B$:
\begin{multline*}
\byz(B) \defeq \{p | p \in \eqvc(B)\\
{} \lor \exists b \in B \cdot \node(b) = p \land  (\lnot \wf(b) \lor \lnot \valid(b))\}.
\end{multline*}
For a block $b$, or set of blocks $S$, we use $\byz(b)$ as a shorthand for $\byz({\preceq}b)$ and $\byz(S)$ for $\byz({\preceq}S)$.

A blocklace storing a CRDT state can be viewed as a pure operation-based CRDT: a subset of blocks will define the PO-Log of updates, with partial order given by $\prec$. Different nodes may have different blocklaces, but the inclusion of a block in the PO-Log only depends on the closure of the block. Given blocklace $B$, the PO-Log is given by:
\begin{multline*}
\polog(B) \defeq {} \\
(\{ b \in B | \wf(b) \land \valid(b) \land \node(b) \notin \byz(b)\}, \prec). 
\end{multline*}
Criteria for validity of an update, specified by the data type, depend on updates in its causal past, as given by the data type state. Having the PO-Log as a subset of the blocklace means that the predicate $\valid$ should only make use of blocks that are part of the state. This prompts redefining block validity, for a block $b = (i, (v, P))$ as:
\[
\valid(b) \defeq \valid(v, \polog({\prec}b)),
\]
using the $\valid$ predicate, but just over the subset of preceding blocks that are part of the PO-Log.  While this results in mutually recursive definitions, the recursion is well-founded, as the functions recur over smaller (prefix) blocklaces.

\section{Byzantine-Repelling Blocklace and Protocol}

Here we extend the CRDT protocol so that all correct nodes eventually exclude (repel) new blocks by Byzantine nodes while still achieving eventual visibility and convergence.

To achieve that we assume that correct nodes satisfy the following Axiom:
\begin{definition}[Node Liveness Axiom]\label{definition:liveness-axiom}
A correct node $p$:
\begin{enumerate}
    \item \textbf{Production:} Produces an initial $p$-block and produces new $p$-blocks indefinitely.
    \item \textbf{Dissemination:}  Eventually sends each block in the closure of every $p$-block to every correct node.
\end{enumerate}
\end{definition}
The term `eventually' in the Dissemination requirement leaves a great deal of freedom to the actual dissemination protocol: It could be arbitrarily lazy as well as probabilistic.
Note that together with a Network Reliability assumption that a message sent among correct nodes 
is eventually received, the Node Liveness Axiom implies the Dissemination Axiom (Definition \ref{definition:dissemination-axiom}).

\subsection{Background and concept}
Due to network asynchrony, different nodes may observe two equivocating blocks in a different order,  so excluding the second block of an equivocation may violate eventual visibility. Furthermore, an equivocator may create more than two equivocating blocks, so different nodes may initially observe different equivocations by the same node.




To limit the harm caused by equivocators, we enhance the use of the buffer $D$ so that a node $p$ that has evidence of other nodes being Byzantine may delay their blocks, as well as delay blocks from other nodes until they acknowledge Byzantine behavior.  For such delays not to result in deadlock, a correct node adheres to the following principles:
\begin{enumerate}[leftmargin=*,topsep=0pt]
    \item A $q$-block $b$ by a correct node $q$ is eventually accepted.
    \item A $q$-block $b$ by a node $q$ is eventually accepted if 
     $b$ provides the first evidence of $q$ being Byzantine.
    \item Block $b$ is accepted if required to satisfy one of the first two principles.
\end{enumerate}

Adhering to these principles is made complicated by colluders.
A node $p$ may \emph{collude} with a Byzantine node  $q$,
creating blocks that observe some $q$-blocks but selectively ignoring others,
never observing a block that is ill-formed, invalid, or provides evidence for an equivocation by $q$.
So, $p$ may never create a block that acknowledges $q$ as Byzantine.

\begin{definition}[Colludes]
A node $p$ colludes with a Byzantine node $q$ if  for
any $p$-block $b$, $q \not\in \byz(b)$.
\end{definition}

To be a colluder, $p$ must violate liveness, as it must refuse indefinitely to accept blocks by correct nodes that acknowledge $q$ as Byzantine.  As such, $p$ is also Byzantine, alas one that can be exposed only at infinity, since at any finite point in time the behavior of $p$ may be attributed to $p$  not having yet received blocks that implicate $q$ as Byzantine, due to network asynchrony.

Thus, by not incurring any behavior that would expose itself as Byzantine,  a colluder $p$ can remain undetected indefinitely, allowing it to create any number of blocks, preceded by any number of blocks from the Byzantine node $q$, indefinitely, and thus polluting the blocklace of any node accepting $p$-blocks by arbitrarily-many Byzantine $q$-blocks.

\subsection{Byzantine-repelling blocklace}

Our approach to address this problem is for each node to repel blocks by Byzantine nodes once detected as such, and expect other correct nodes to do the same. The joint behavior of correct nodes eventually repel any new blocks by any Byzantine node, even in the presence of colluders.



The challenge is that for a node $q$ known to be Byzantine by a correct node $p$, node $p$
would never be able to prove that some node $r$ is a colluder with $q$. What $p$ can do is
buffer $r$-blocks until (if ever) it receives an $r$-block $b$ with $q \in \byz(b)$, proving that $r$ is not colluding with $q$ by recognizing $q$ as Byzantine.
A correct node $r$ will eventually do that, and so $p$ can wait for such an $r$-block
$b$ to arrive, upon which it can incorporate $b$ and its closure.

Assume $q$ is the only Byzantine node known to $p$ at some point of the computation and that $r$ aims to collude  with $q$. Then $r$-blocks will be buffered by $p$ and not accepted, except perhaps for some $r$-block $b'$ that precedes some block $b$ from a possibly-correct node $s$ that proves itself not to collude with $q$, with $b'$ created before $s$ knew $q$ to be Byzantine.
Eventually, all correct nodes will know $q$ to be Byzantine and will
repel not only $q$-blocks but also blocks from colluders with $q$.

To realize the principles above and ensure that harm caused by Byzantine nodes is bounded, even in the presence of colluders, we employ the following definition of a Byzantine-repelling blocklace:

\begin{definition}[Byzantine-Repelling Blocklace]
A blocklace $B$ is \emph{Byzantine-repelling}, written $\brep(B)$, if either $B = \emptyset$ or there exists a prefix blocklace $B' \subset B$, with $B = B' \union {\preceq} b$ for some $b \in \max(B)$, such that $\brep(B')$ and either:
\begin{enumerate}
    \item $\byz(B \setminus \{b\}) \subset \byz(B)$, or
    \item $\node(b) \notin \byz(B) \land \byz(B') \subseteq \byz(b)$.
\end{enumerate}
We use $\brep(b)$ as a shorthand for $\brep({\preceq}b)$.
\end{definition}

Correct nodes will always maintain Byzantine-repelling blocklaces, either when incorporating blocks from other nodes or generating new blocks. This invariant can be used to enhance the detection of Byzantine nodes. If the blocklace $B$ that precedes a new block is not Byzantine-repelling, then that node is Byzantine. This allows redefining $\byz$ as:
\begin{multline*}
\byz(B) \defeq \{p | p \in \eqvc(B) \lor \exists b \in B \cdot \node(b) = p \\
{} \land (\lnot \wf(b) \lor \lnot \valid(b) \lor \lnot \brep({\prec}b))\}.
\end{multline*}
This results in a mutually recursive definition, which, as before, remains well-founded.


\subsection{Byzantine-repelling protocol}

The Byzantine-repelling protocol uses the definition in the following procedure for a node accepting received blocks from the buffer $D$ into the blocklace $B$:  Incorporate $S = {\preceq}b \cap D$ at once, for some $b \in D$, resulting in $D' = D \setminus S$ and
$B' = B \union S$, when ${\preceq}S \subseteq B \union S$ and provided the resulting blocklace $B'$ is Byzantine-repelling. This can be checked directly from the definition, with the roles of $B$ and $B'$ reversed. The current Bytantine-repelling blocklace $B$ is the existentially assumed prefix and $b$ is the assumed maximal in the resulting blocklace $B'$.

Note that if a node $p$ follows this protocol then every $p$-block is Byzantine-repelling.  Contrapositively, a $p$-block that is not Byzantine-repelling is evidence for $p$ not following the protocol, namely being Byzantine.  Thus, in the context of this protocol, the notion of Byzantine nodes is extended, as shown above, to include nodes producing blocks that are not Byzantine-repelling.


We claim that this procedure ensures the following Propositions, proven in Appendix \ref{appendix:proofs}.

\begin{proposition}[Eventual Visibility]\label{proposition:eventualy-visibility}
Any block in a blocklace of a correct node is eventually in the blocklace of every correct node.
\end{proposition}

We say an a block is \emph{public} if  it is known to at least one correct node.   An \emph{evidence} for a node $p$ being Byzantine is a malformed or invalid  $p$-block or two $p$-blocks that constitute an equivocation.

\begin{proposition}[Finite Harm]\label{proposition: finite-harm}
If $p$ is correct and there is public evidence that $q$ is Byzantine then $p$ will eventually stop including $q$-blocks in its blocklace.
\end{proposition}

%% file: appendix.tex

\appendix{}

\section{Proofs}\label{appendix:proofs}.

To prove  Proposition \ref{proposition:eventualy-visibility} we need the following Lemma:
\begin{lemma}[Byzantine Convergence]\label{lemma:Byzantine-convergence}
There is a time, referred to as a \temph{Byzantine Convergence Time (BCT)}, beyond which $\byz(b)=\byz(b')$ for every new $p$-block $b$ and new $q$-block $b'$, upon which we refer to the set $\byz(b)$ as the \temph{convergent set of Byzantines}.
\end{lemma}
\begin{proof}
By way of contradiction and wlog assume there is a node $r$ such that $r\in \byz(b)$ for some $p$-block $b$, but $r\notin \byz(b')$ for any $q$-block $b'$.
By the Node Liveness Axiom, $p$ will eventually send ${\preceq}b$ to $q$, including $r$-blocks that constitute evidence for $r$ being Byzantine.  By definition of the protocol, since $r$ would be a new Byzantine to $q$, $q$ would accept $r$-blocks providing such evidence into its blocklace $B$.  By Node Liveness, $q$ would eventually produce a new block $b'$ that includes such evidence in its closure, namely  $r\in \byz(b')$, a contradiction.
\end{proof}


%
%

\begin{proof}[Proof of Proposition \ref{proposition:eventualy-visibility}]
Let $b$ be any block in the blocklace of a correct node $p$ and consider another correct node $q$.
Consider a $p$-block $b_p$ produced when $b$ is present, which will eventually occur due to the Node Liveness Axiom, and a $q$-block $b_q$, with both $b_p$ and $b_q$ produced after the Byzantine convergence Time.
Since $p$ and $q$ are correct by assumption, by Lemma \ref{lemma:Byzantine-convergence} $\byz(b_p) = \byz(b_q)$. Also by Node Liveness, ${\preceq}b_p$, which includes $b$, will be eventually received by $q$, and by definition of the protocol it will be incorporated in the blocklace of $q$.
\end{proof}

\begin{proof}[Proof of Proposition \ref{proposition: finite-harm}]
Due to the Dissemination Axiom and the Byzantine-Repelling protocol, eventually the blocklace $B$ of $p$ will include evidence that $q$ is Byzantine, namely $q\in \byz(B)$, and a $p$-block $b$ will be produced at with $q \in \byz(b)$.
At that point,  let $R_q(B)$ be the set of nodes such that for each $r\in R$, (\ia)  $r\notin \byz(B)$ and (\ib) $B$ does not include an $r$-block $b'$ such that $q\in \byz(b')$.

According to the protocol, $p$ will add a new $q$-block $b$ to $B$ only if $b$ precedes an $r$-block $b'$ for some $r\in R_q(B)$ such that $q \in \byz(b')$.  Hence, after $b$ is added to $B$, $r$ is removed from $R_q(B)$ by its definition, and no more $r$-blocks pointing directly to a $q$-block can be added, and the size of $R_q(B)$ decreases by $1$.

Every correct node in $R_q(B)$ will eventually receive the evidence that $q$ is Byzantine, by  Node Liveness will produce a block acknowledging this evidence,  and will be removed  from $R_q(B)$.
Similarly, any node node in $R_q(B)$ that produces public evidence for being Byzantine will also be eventually removed from $R_q(B)$.  This leaves $R_q(B)$ with only possible colluders with $q$; any such 
colluder $q'$ will not provide evidence that incriminates itself as Byzantine, nor will it create a block $b''$ such that $q\in \byz(b'')$.
Hence, further blocks by $q'$, as well as any preceding $q$-blocks not already in $B$, will never be incorporated in $B$, as $q'$ will never acknowledge that $q$ is Byzantine.

Since $R_q(B)$ was finite to begin with, removal of nodes from it can be repeated only a finite number of times,  hence eventually $R(B)$ will include only colluders, if any, upon which no more $q$-blocks can be added to $B$. 
\end{proof}